\documentclass[aip,apl,twocolumn,reprint]{revtex4-1}
\usepackage{graphicx}%
\usepackage{dcolumn}%
\usepackage{bm}
\begin{document}
\title{ Sub-micron normal-metal/insulator/superconductor tunnel junction thermometer and cooler using Nb}
\author{M. R. Nevala}
\author{S. Chaudhuri}
\author{J. Halkosaari}
\author{J. T. Karvonen}
\author{I. J. Maasilta} 
\affiliation{Nanoscience Center, Department of Physics, P. O. Box 35, FI-40014 University of Jyv\"askyl\"a, Finland}

\begin{abstract}
We have successfully fabricated  Cu/AlOx-Al/Nb normal-metal/insulator/superconductor
tunnel junction devices with a high value of the superconducting gap (up to $\sim 1$ mV), using electron-beam lithography
and angle evaporation techniques in the sub-micron scale. The subgap conductance of these   junctions shows the expected strong
temperature dependence, rendering them suitable for thermometry all the way from 100 mK to 6 K. In addition, some direct electronic
cooling of the normal metal was also seen at bias values near the gap edge. The device performance was strongly influenced by the details of the Al layer geometry, with  lateral spilling of the aluminium giving rise to strong extra subgap features, and the thickness of Al layer affecting the proximised superconducting gap value of the superconducting Al/Nb bilayer.

\end{abstract}
\maketitle


Normal-metal-insulator-superconductor (NIS) tunnel junction devices have already proven as promising solid state coolers \cite{muhonen,RevModPhys.78.217,PhysRevLett.102.165502,NISTrecent}, accurate low temperature thermometers and sensitive bolometers \cite{RevModPhys.78.217, Koppinen, NISTbolo}, and near-ideal single electron transistors\cite{Pekolanatphys,PhysRevB.85.012504}. Some active research on such devices is aimed at enhancing the cooling
power at sub-300 mK temperatures\cite{NISTnew}, but increasing the operational range in temperature by changing the superconducting material (gap $\Delta$) from the traditional aluminum has not been very successful yet, although a promising new alternative is to use SIS' structures, instead \cite{giazotto}. 

 To extend the range of NIS devices beyond the operational range of Al, which has an upper limit for thermometry at $T_{C}\sim$1.5 K and the maximum cooling power at $T \sim $ 300 mK, one should consider the elemental metal with highest gap, niobium ($T_C $ $\sim$ 9 K). It is already routinely used for SIS tunnel junction applications  in SQUIDs, radiation detectors and digital electronics \cite{vanDuzer}. In those applications, fairly large micron-scale junctions can typically be tolerated, and fabrication usually proceeds by the robust Nb/Al/AlOx/Nb trilayer deposition and etching techniques \cite{gurv,tolpygo}, where Nb and Al are typically sputter deposited, and the high quality AlOx barrier is thermally grown on a thin $<$ 10 nm Al layer. This process typically yields a high gap $\Delta \sim$ 1.3 mV and a near bulk $T_{C} \sim$ 9 K \cite{morohashi,dolatajap}. 

In contrast, for single-charge devices and for small thermometers and bolometers, sub-micron scale junctions are desired. They are easier to fabricate using angle-evaporation and lift-off \cite{dolan}, although a successful but more complex sub-micron Nb trilayer Josephson junction process has also been demonstrated \cite{dolatajap}. The quality of evaporated Nb, unfortunately, has turned out to be quite sensitive to the exact chamber and substrate conditions, such as vacuum level, evaporation speed, substrate to Nb crucible distance and especially to the type of resist used \cite{Kim,Pekola1999653,Dolata,Hadara,Dubos,Hoss}. Problems with standard polymer resists are typically attributed to decomposition and outgassing during Nb evaporation, leading to the suppression of $T_C$ and $\Delta$.  
 
Here, we demonstrate that a fairly simple angle-evaporation process can, nevertheless,  be used to fabricate good quality micron to submicron scale NIS tunnel junctions with Al/Nb bilayer as the superconductor and Cu as the normal metal, and that the junctions can be used for sensitive thermometry up to 6 K, with a full theoretical understanding of the response. Electronic cooling was also demonstrated near the gap edge at an elevated temperature range compared to standard Al coolers.  The only previous work on Cu/AlOx/Nb NIS junctions used a trilayer technique to produce only large-scale ($ \sim 100 \mu$m) junctions, and reported non-ideal responsivity with no cooling \cite{castellano,leoni}.

A scanning electron micrograph of the  Nb/Al-AlO$_{x}$/Cu multiple NIS junction sample 
fabricated using electron beam lithography (EBL) and
shadow-evaporation technique is shown in Figure 1(a), with a close-up of the junction area in Fig. 1 (b). Oxidized silicon serves as  the substrate.  A dual layer PMMA/P(MMA--MAA) positive resist recipe was used for the lithography, forming an undercut structure \cite{note}. 
The metal evaporation was performed in an ultra high vacuum chamber with an electron-beam source. First, 20 nm of Nb was
evaporated from an angle of 30 degrees with respect to the plane of the substrate. After the deposition of Nb,
we waited 15 minutes for the Nb to cool down. This leads to an improvement in the quality of the junction\cite{Simon}. This was followed by
Al evaporation (5-20 nm thickness), whereby half of the total Al to be deposited was evaporated from the same 30 degree angle as that of Nb, while the other half was
deposited from -30 degrees.  The tip of the Nb electrode was designed to be slightly
tapering, as shown in Figure 1(b), to improve Al coverage. The idea of the Al deposition from two opposite angles was to achieve uniform
Al coverage, even on the edges of the Nb electrode. The Al was then oxidized in-situ at 350 mbar for 25 minutes. Finally, 30-50 nm copper
was evaporated from the normal angle. The chamber pressure during the metal deposition was 1.5$\times$10$^{-8}$ mbar, and the evaporation rates
for Nb, Al and Cu were typically 0.45, 0.1 and 0.1 nm/s, respectively.

We have succesfully made samples with four different  thicknesses ($t_{Al,Cu}$) of  Al (Cu) layers, namely 7 (35), 10 (40), 15 (30) and 20 (50) nm, while Nb was 20 nm thick
in all cases. An attempt was made to use a 5 nm Al layer, as well, but the Al coverage was not complete anymore in that case. The thickness and the line width ($ \sim $1 $\mu$m) for the Nb electrodes were chosen so that on one hand, they could still show
$T_{C}$ and $\Delta$ close to the bulk values, while on the other hand, it would still be possible to cover the Nb layer with a thin film of Al.  The dimensions of the Cu wire were typically  55 $\mu$ m $\times$ 400 nm  $\times$ 35 nm. We have also confirmed the earlier observations that $T_{C}$ and $\Delta$ of shadow-evaporated Nb depend strongly of the Nb thickness, line width and deposition conditions \cite{Pekola1999653,Kim,Dolata,Hadara,Dubos,Hoss,PhysRevLett.95.147003,PhysRevLett.57.901}.

\begin{figure}[t]
\includegraphics[width=0.35\columnwidth]{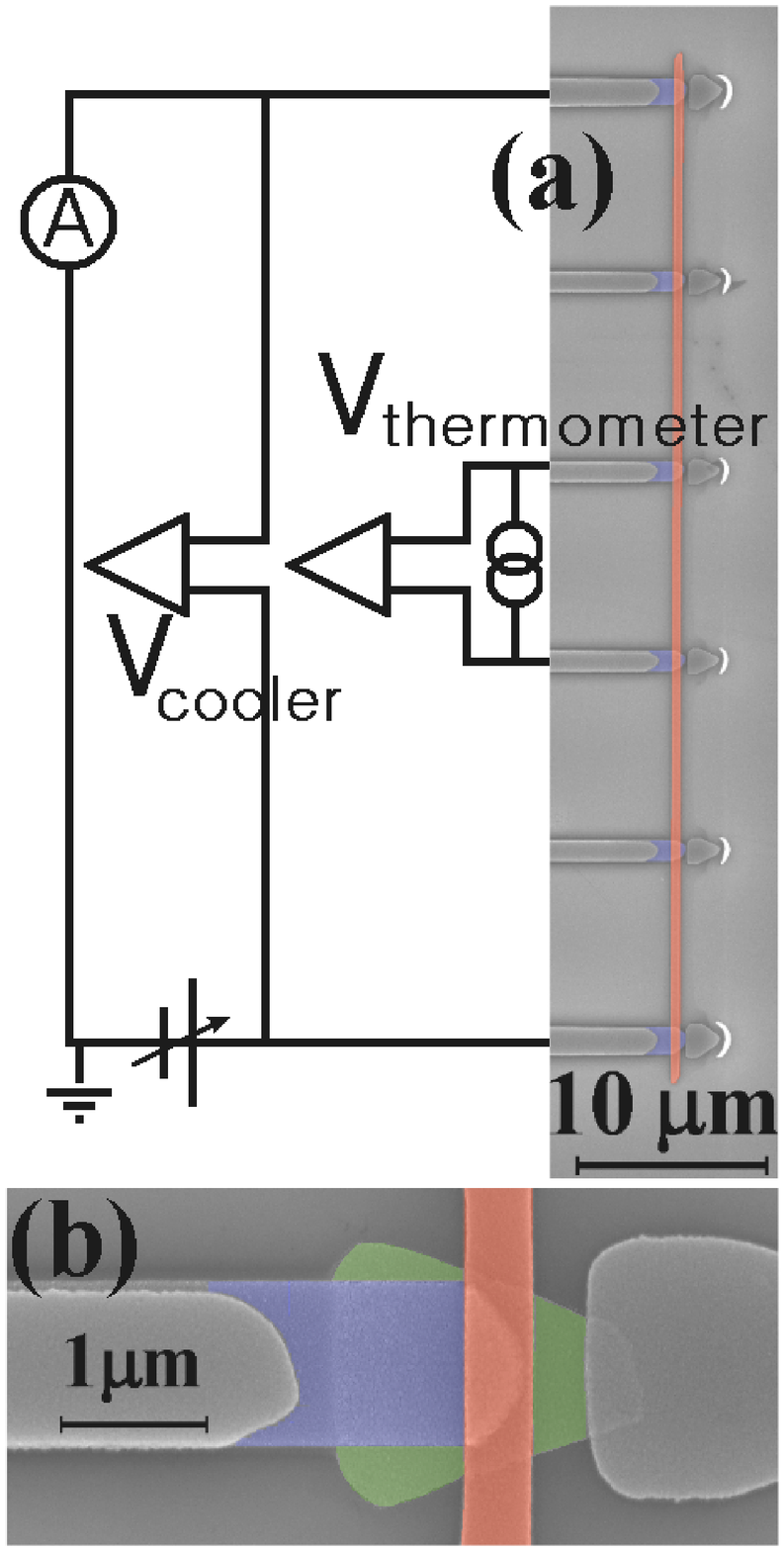}
\includegraphics[width=0.6\columnwidth]{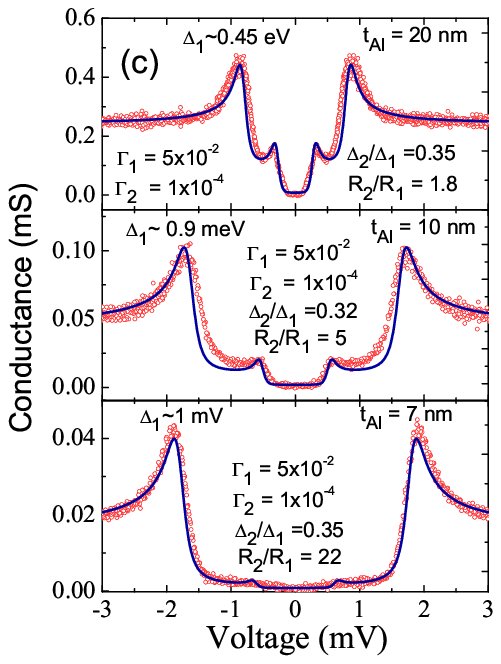}
\caption{(Color online) (a) Scanning electron micrograph (SEM) of a sample  showing the measurement configuration. The outermost two larger electrodes serve as the cooler junctions, while a pair of inner electrodes act as a SINIS thermometer to sense the electron temperature of the Cu wire.  (b) High resolution  SEM of a single junction showing  the tapered Nb edge (blue) covered with Al (green), contacting the Cu (red). (c) Measured $dI/dV-V$ characteristics of a SINIS junction pair, indicated by the red dots, at a bath temperature of 150 mK,  for samples with three different thicknesses of the Al layer (7, 10, 20 nm). Theoretical fits using a parallel junction model is represented by the lines, with parameters shown in the figure.}
\end{figure}
The measurements
were carried out using a He$^3$-He$^4$ dilution refrigerator with a base temperature of 60 mK.  The measurement lines had two RC
filter stages, one at 4 K and the other at 60 mK,  and microwave filtering  between the RC filters was achieved with the
help of Thermocoax cables connecting the two RC filter stages. Figure 1(c) shows the differental conductance-voltage ($dI/dV-V$) characteristics, measured at a bath temperature of 150 mK,  for samples with  three different thicknesses (7, 10 and 20 nm) of the Al layer. Clearly, the characteristics show a double gapped structure (two sets of conductance peaks), which evolves with the Al thickness.
This can be explained as follows:  As there is a slight overspill of Al on top of the Nb to ensure complete coverage of the Nb (avoiding a direct Nb-Cu contact), 
part of the current  tunnels into Cu through the overspill region where Al is not fully proximised, as the overspill length can be of the same order of magnitude or larger than the coherence length. Thus, each physical Nb-Al-AlO$_{x}$-Cu junction can be modelled as  two electrical junctions in parallel, the main one (J$_{1}$) with Al directly on top of Nb, and the other (J$_{2}$) the overspill region without Nb. The gap value of the first junction ($\Delta_1$) is that of Al proximised by Nb in the perpendicular direction, whereas  the gap value of the overspill junction ($\Delta_2$) is that of Al proximised by Nb in the lateral (in-plane) direction, which in the case of large overspill would lead to an unproximized Al gap.  

The total current as a function of voltage $V$ can then be written as a sum of two junction currents, where each individual current $I_{J_i}$ is given by the usual expression \cite{RevModPhys.78.217} $I_{J_i}=\frac{1}{eR_{i}}\int_{-\infty}^{\infty}\!\! d\epsilon N_{S,i}(\epsilon )[f_{S}(\epsilon )-f_{N}(\epsilon+eV )]$, where $R_{i}$ is the tunneling resistance of junction $i$, $f_{N/S}(\epsilon)$ is the Fermi function in either the superconducting or the normal electrode, and $N_{S,i}(\epsilon)$ is the normalized superconducting quasiparticle density of states of junction $i$. Here, we use the expression $N_{S,i}(\epsilon, T_{S})  =\left | {\rm Re} \left \{ (\epsilon+i\Gamma_i)/\sqrt{(\epsilon+i\Gamma_i)^{2}-\Delta_{i}^{2}(T _{S} )} \right \} \right |$, which for Al junctions has been shown \cite{PhysRevLett.53.2437,Pekoladynes} to accurately describe the broadening  of the DOS singularity,  where $ \Gamma_i$  is the Dynes parameter  describing the broadening, and $\Delta_i(T_{S})$  is the superconducting energy gap, with its temperature dependence written explicitly. Intuitively,  it is clear that this model produces peak features at two voltages $eV = \Delta_{1}$ and $eV=\Delta_{2}$,  but the specific shape of the conductance curve is determined by the ratio of the two gaps ($\Delta_{2}/\Delta_{1}$) and the ratio of the tunnelling resistances ($R_{2}/R_{1}$). 
\begin{figure}[t]
\includegraphics[width=1\columnwidth]{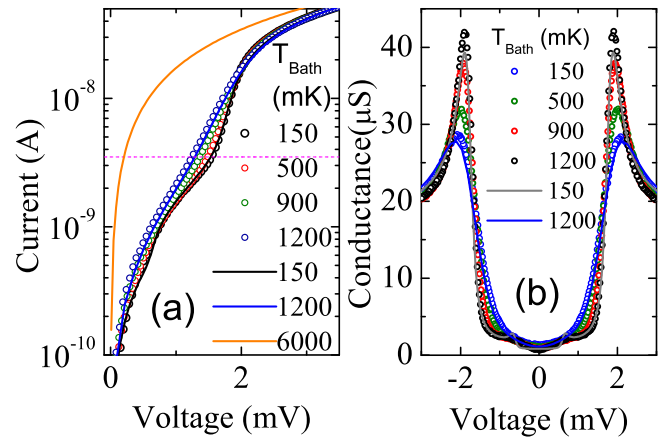}
\caption{(Color online) Bath temperature dependence of (a) $I-V$  and  (b)  $dI/dV-V$ characteristics of a SINIS pair with  7 nm thick aluminium, between 0.15 - 1.2 K. The dots are the experimental data while the lines are theoretical fits for $ T_{bath } $= 150 mK, 1.2 K and 6 K (normal state). The dotted line in  (a) represent a constant current bias of 3.5 nA.}
\end{figure}

We have succesfully used the parallel junction model to fit the conductance plots of Figure 1 (c). The larger gap $\Delta_{1}$ (perpendicularly proximized Al/Nb) increases with decreasing $t_{Al} $ from 0.45 and 0.9 up to 1 mV for $ t_{Al} $ =  20, 10 and 7 nm, respectively. To our knowledge, 1 mV is the highest ever gap reported for a tunnel junction with a Nb electrode fabricated by angle evaporation \cite{Dolata,Kim,Hadara}. From the fits in Figure 1(c), we also find that $\Delta_{2}/\Delta_{1}\sim $ 0.30-0.35 for all samples. On the other hand, the ratio of the tunneling resistances $R_{2}/R_{1}$ is not constant, but varies by an order of magnitude between $\sim$ 2-20, reflecting the different relative strengths of the two conductance peaks. Physically, this results from differing amounts of Al overspill between the three samples. The Al overspill for the samples with 10 and  20 nm thick Al layer  is about 200-300 nm. This makes the two junction areas and thus $R_{2}$ and $R_{1}$ comparable, leading to clear exhibition of the small gap feature in Fig. 1 (c).  For the  sample with 7 nm thick Al, we managed to reduce the lateral spill to $<$ 100 nm by improving the Al to Nb layer alignment, and were thus able to suppress the small gap features significantly. The  measured $T_{C}$  of the sample  with $ t_{Al} $ = 7 nm with  was $ \sim $ 6 K, leading to a relation $\Delta_{1} \approx 1.9k_{B}T_{C}$, in agreement with the literature for pure Nb \cite{vanDuzer}. Finally, the relative Dynes parameter values $\Gamma_i/\Delta_i$  stay approximately constant regardless of the Al thickness with values $\Gamma_{1}/\Delta_{1} \sim 5 \times 10^{-2}$, and $\Gamma_{2}/\Delta_{2} \sim 1 \times 10^{-4}$. $\Gamma_{2}$ is consistent with previous studies in Al \cite{Koppinen,Pekoladynes,Greibe}, but $\Gamma_{2}$, interestingly, is smaller than previously reported for pure Nb \cite{proslier}. 

The temperature dependence of the current-voltage and conductance-voltage characteristics of the sample with 7 nm Al  are shown in Fig. 2, with other samples showing similar behavior (not shown). The low bias region $ < 1$ mV is dominated by the excess sub-gap current of the $\Delta_1$ gap edge (parametrized by $\Gamma_1$), and has a weak temperature dependence. However, at higher bias voltages, a clear temperature dependence is seen, as expected for ideal NIS junctions \cite{RevModPhys.78.217}. The theoretical curves based on the parallel junction model fit the data well at all temperatures, if self-cooling, heating and electron-phonon (e-p) coupling in the Cu wire are taken into account \cite{chaudhuriPRB}.  

\begin{figure}[t]
\includegraphics[width=1\columnwidth]{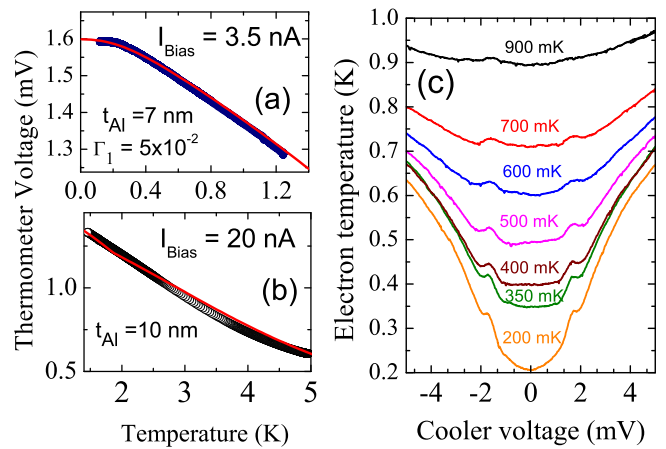}
\caption{(Color online) Thermometry characteristics for  the sample with (a) 7 nm Al, bias current $\sim$ 3.5  nA, and (b) 10 nm Al, bias current 20 nA. The solid lines indicate the corresponding theoretical fits.  (c) Measured electron temperature of the copper wire as a function of cooler voltage at several bath temperatures. }
\end{figure}
As can be seen from Fig. 2, the junctions can be used for thermometry by biasing with constant current above the excess sub-gap current level [$> 3$ nA in Fig. 2 (a)], and by measuring the temperature dependent voltage. The response of two of the devices ($ t_{Al} $ = 7 and 10 nm) to changes in bath temperature $ T_{bath} $ are shown in Figs. 3(a) and (b),  along with the corresponding theoretical fits, assuming that the electron temperature is equal to the bath temperature (no self-cooling, heating and e-p coupling). Clearly, our devices have good responsivity $dV/dT \sim $ 0.2-0.3 mV/K all the way from 200 mK to $ \sim $ 5 K, and some responsivity left up to $T_C \sim 6$ K (not shown).  We also see that the fits are good for these bias points even without the thermal effects taken into account, so that we have full understanding of the temperature response of the devices. The responsivity is the same as in Al-based NIS thermometers as expected from the theory, but the major advantage of Nb-based NIS thermometry is that the operational range is extended to much higher temperatures. 

In addition to thermometry, we have investigated the electronic cooling performance  of the device with $ t_{Al} $ = 7 nm, shown in Figure 3(c), by measuring the temperature of the Cu island with one SINIS pair as a function of the bias voltage across a second (cooler) SINIS pair [the measurement circuit shown in Fig. 1 (a)]. A clear cooling dip is observable at all values of $T_{bath}=200$ mK - 900 mK near the second gap edge at $V \sim 2$ mV. Unfortunately, there is also observable heating in the subgap region (especially at low $T$), caused by the excess subgap currents, so that no cooling below the base temperature was obtained yet. However, the observed value of $\Gamma_1$ does not theoretically prevent reaching temperatures below $ T_{bath} $, and as this was the first demonstration of Nb-based NIS cooling, we  believe that there is room for improvement. For example, the junction size and geometry could be improved, the overspill could be reduced even further, and the normal metal volume could be reduced by at least an order of magnitude. 

In conclusion, we have successfully fabricated submicron scale Nb-Al-AlO$_{x}$-Cu based NIS tunnel junctions using e-beam lithography and angle evaporation, with a gap $\Delta \sim $ 1 mV and broadening parameter $\Gamma/\Delta \sim 5 \times 10^{-2}$.
The devices have good thermometric response in the temperature range between 200 mK to 5 K. Direct electronic cooling was observed at all bath temperatures between 200 mK and 900 mK, but not below the base temperature yet.

This research was supported by the Academy of Finland project number 128532. 


\begin{thebibliography}{33}
\bibitem [{not()}]{note}%
  \BibitemOpen
  \href@noop {} {\bibinfo  {journal} {For the co-polymer layer, we have used
  commercially available P(MMA--MAA) resist diluted in acetic acid, while the
  top layer resist is PMMA diluted with anisole. The layer thicknesses were 500
  and 450 nm, respectively, while the respective baking times were 15 and 5
  minutes at a temperature of $170^\circ$ C. Before metal evaporation, the
  sample was cleaned in oxygen plasma for 30 seconds to get rid of organic
  impurities on the exposed substrate surfaces}\ }\BibitemShut {NoStop}%
\end{thebibliography}

%

\end{document}